
\input phyzzx

%
\catcode`\@=11 
\def\papers{\papersize\headline=\paperheadline\footline=\paperfootline}
\def\papersize{\hsize=40pc \vsize=53pc \hoffset=0pc \voffset=1pc
   \advance\hoffset by\HOFFSET \advance\voffset by\VOFFSET
   \pagebottomfiller=0pc
   \skip\footins=\bigskipamount \normalspace }
\catcode`\@=12 
\papers

\tolerance=500000
\overfullrule=0pt

\pubnum={PUPT-1379  \cr
hep-th@xxx/9301119}

\date={January 1993}
\pubtype={}
\titlepage
\title{HAMILTONIAN APPROACH TO
2D DILATON-GRAVITIES\break
 AND INVARIANT ADM MASS}
\author{{
Adel~Bilal}\foot{
 on leave of absence from
Laboratoire de Physique Th\'eorique de l'Ecole
Normale Sup\'erieure, \nextline 24 rue Lhomond, 75231
Paris Cedex 05, France
(unit\'e propre du CNRS)\nextline
e-mail: bilal@puhep1.princeton.edu
}}
\andauthor{{
Ian I. Kogan}\foot{
 on leave of absence from ITEP,  B. Cheremyshkinskaya
 25, Moscow 117259,
Russia \nextline
e-mail: kogan@puhep1.princeton.edu
}}
\address{\it Joseph Henry Laboratories\break
Princeton University\break Princeton, NJ
08544, USA}

\vskip 3.mm
\abstract{The formula existing in the literature for the
ADM mass of 2D dilaton gravity is incomplete. For example,
in the case of an infalling matter shockwave this formula
fails to give a time-independent mass, unless a very
special coordinate system is chosen. We carefully carry out
the canonical formulation of 2D dilaton gravity theories
(classical, CGHS and RST). As in 4D general relativity one
must add a boundary term to the bulk Hamiltonian to obtain
a well-defined variational problem. This boundary term
coincides with the numerical value of the Hamiltonian and
gives the correct mass which obviously is time-independent.
}

\endpage
\pagenumber=1

 \def\PL #1 #2 #3 {Phys.~Lett.~{\bf #1} (#2) #3}
 \def\NP #1 #2 #3 {Nucl.~Phys.~{\bf #1} (#2) #3}
 \def\PR #1 #2 #3 {Phys.~Rev.~{\bf #1} (#2) #3}
 \def\PRL #1 #2 #3 {Phys.~Rev.~Lett.~{\bf #1} (#2) #3}
 \def\CMP #1 #2 #3 {Comm.~Math.~Phys.~{\bf #1} (#2) #3}
 \def\IJMP #1 #2 #3 {Int.~J.~Mod.~Phys.~{\bf #1} (#2) #3}
 \def\JETP #1 #2 #3 {Sov.~Phys.~JETP.~{\bf #1} (#2) #3}
 \def\PRS #1 #2 #3 {Proc.~Roy.~Soc.~{\bf #1} (#2) #3}
 \def\IM #1 #2 #3 {Inv.~Math.~{\bf #1} (#2) #3}
 \def\JFA #1 #2 #3 {J.~Funkt.~Anal.~{\bf #1} (#2) #3}
 \def\LMP #1 #2 #3 {Lett.~Math.~Phys.~{\bf #1} (#2) #3}
 \def\IJMP #1 #2 #3 {Int.~J.~Mod.~Phys.~{\bf #1} (#2) #3}
 \def\FAA #1 #2 #3 {Funct.~Anal.~Appl.~{\bf #1} (#2) #3}
 \def\AP #1 #2 #3 {Ann.~Phys.~{\bf #1} (#2) #3}
 \def\MPL #1 #2 #3 {Mod.~Phys.~Lett.~{\bf #1} (#2) #3}

\def\d{\delta}
\def\pa{\partial}

\def\Scl{S_{\rm cl}}

\def\f{\phi}
\def\r{\rho}
\def\o{\omega}

\def\s{\sigma}

\def\l{\lambda}
\def\t{\tau}
\def\st{\tilde\sigma}
\def\tt{\tilde\tau}

\def\is{\int {\rm d}^2 \sigma}
\def\ds{{\rm d} \sigma}
\def\rg{\sqrt{-g}}
\def\k{\kappa}

\def\m{\mu}
\def\n{\nu}

\def\zt{{\tilde Z}}

\def\NS{ {/\kern -0.60em \nabla } }

\def\stp{\s\to\infty}
\def\stm{\s\to -\infty}
\def\sep{\s =+\infty}
\def\sem{\s =-\infty}
\def\pf{\Pi_\f}
\def\pr{\Pi_\r}
\def\pz{\Pi_Z}
\def\ca{{\cal C}_A}
\def\cb{{\cal C}_B}

{ \chapter{Introduction}}

The classical action for dilaton gravity in two dimensions
is
 \REF\CGHS{C. Callan, S. Giddings, J. Harvey and A.
Strominger, \PR D45 1992 R1005 .} [\CGHS]
$$
\Scl = {1\over 4\pi} \is \rg \left[ e^{-2\f}\left(
R+4(\nabla\f)^2 + 4\l^2\right) -\half \sum_{i=1}^N
(\nabla f_i)^2\right]
\eqn\i$$
where $\f$ is the dilaton and $f_i$ are $N$ matter
fields. The virtues of this action and its quantum versions
have been discussed in the literature \REF\BC{A. Bilal and
C. Callan, ``Liouville models for black hole evaporation",
Princeton University preprint PUPT-1320, to appear in Nucl.
Phys. B., hep-th@xxx/9205089.}  \REF\ALW{S. de Alwis,
\PL B289 1992 278 , ``Black hole
physics from Liouville theory", Boulder preprint
COLO-HEP-284, hep-th@xxx/9206020.}  \REF\RST{J. Russo, L.
Susskind and L. Thorlacius, \PR D46   1992  3444 .}
\REF\CC{J. Russo, L.
Susskind and L. Thorlacius,
 ``Cosmic censorship in two-dimensional gravity", Stanford
preprint SU-ITP-92-24, hep-th@xxx/9209012.}
[\CGHS - \CC] (see also
\REF\CHAM{A. Chamseddine, \PL B256 1991 379 , \NP B368 1992
98 .}
ref. \CHAM\ for earlier work) and we won't repeat them
here\foot{ See ref. \CGHS\ or \BC\ for all conventions.}
The classical action admits static black hole solutions
$$
{\rm d}s^2 = \left( 1+ {m\over \l}e^{-2\l\s} \right)^{-1}
({\rm d}\s^2 - {\rm d}\t^2 ) \quad , \quad e^{-2\f}=e^{2\l\s}
+ {m\over \l} \ .
\eqn\ii$$
where we use conformal coordinates and the conformal factor
$e^{2\r} $ and the dilaton satisfy $\f =-\l\s +\r$.
Asymptotically, as $\s\to\infty$, the metric becomes
Minkowskian and one approaches the  linear dilaton
vacuum (LDV), $\r=0,\ \f=-\l\s$. The parameter $m$ which
characterizes the asymptotics as $\s\to\infty$, $\r\sim
-{m\over 2\l} e^{-2\l\s},\ \f\sim -\l\s -{m\over
2\l}e^{-2\l\s}$, is the mass of the black hole.

More generally, it was found
\REF\WIT{E. Witten, \PR D44 1991 314 .}
\REF\PS{Y. Park and A.
Strominger,``Positive mass and supersymmetry in classical
and quantum two-dimensional dilaton-gravity", Santa Barbara
preprint UCSBTH-92-39, hep-th@xxx/9210017.}
[\WIT, \CGHS, \PS] that the ADM mass of any configuration
asymptotic to the LDV should be given by
$$M=2\, e^{2\l\s} \left( \pa_\s \d\f+\l\d\r\right)
\Big\vert_{\s=\infty}
\eqn\iii$$
where $\d\f=\f-\f_{\rm LDV},\ \d\r=\r-\r_{\rm LDV}$ are the
deviations from the linear dilaton vacuum. For the static
black hole \ii\ this gives $M=m$ correctly. Thus the mass
is given by the asymptotics of the fields only, as in 4D
general relativity. However, there are two spatial ends of
the world, $\sep$ and $\sem$, and one might
wonder whether one should include some contribution from
$\stm$ into $M$. To circumvent this question for
the moment, we will only consider configurations asymptotic
to the LDV {\it both} for $\s\to +\infty$ and $\stm$. (This
of course excludes \ii.) Then a possible contribution
$2e^{-2\f} \left( \pa_\s \d\f+\l\d\r\right)
\vert_{\sem}$ will vanish. This kind of configurations
are typically encountered if some matter falls into the LDV
($T_{++}\ne 0$ for some finite interval of $\s+\t$). This
produces a black hole which, in the classical case, will not
radiate and should have constant mass equal to the total
energy carried by the infalling matter. If we take
$T_{++}(\s+\t)= a\d(\s+\t)$ for example, the matter carries
total energy $a$ and the solution is (in conformal gauge)
$\f=-\l\s,\ \r=0$ for $\s+\t<0$, while for $\s+\t>0$:
$$\f=-\l\s+\r\ \ ,\ \ \r={a\over 2\l} e^{\l(\t-\s)}
+\left( {a^2\over 4\l^2} e^{2\l\t}-{a\over 2\l}\right)
e^{-2\l\s} +{\cal O}(e^{-3\l\s})\ .
\eqn\iv$$
This is obviously asymptotic to the LDV as $\s\to
\pm\infty$. If we insert this into \iii\ we obtain
$$M=a\left(1-{a\over 2\l}e^{2\l\t}\right)
\eqn\v$$
which depends on time. The origin of this problem can be
traced to the assumption, implicit in the derivation of
\iii, that $\d\f,\ \d\r$ are ${\cal O}(e^{-2\l\s})$, while
\iv\ actually contains terms ${\cal O}(e^{-\l\s})$.

It should be noted that there exists a conformal coordinate
transformation $\st+\tt=\s+\t,\ \st-\tt={1\over
\l}\log\left( e^{\l\s-\l\t}-{a\over\l}\right)$ so that, as
$\st\to\infty$ for fixed $\tt$, one has $\f\sim
-\l\st-{a\over 2\l}e^{-2\l\st},\
 \r\sim -{a\over 2\l}e^{-2\l\st}$. But then $\f$ and $\r$
no longer equal the LDV (by LDV we mean $\f=-\l\st$ and
$\r=0$) for $\s+\t<0$, i.e. for $\st\to -\infty$. In fact
they are not even asymptotic to the LDV. (Making the
transformation only for large $\s+\t$ and patching it
smoothly to the identity for $\s+\t<0$ would obviously not
be conformal.)
Thus, in this case, there exists a special coordinate system,
specified by $\d\r, \d\f ={\cal O}(e^{-2\l\st})$ as
$\st\to\infty$. In this  coordinate system, using \iii, $M=m$
is time-independent, but one has lost the LDV asymptotics
for $\st\to -\infty$. Here we want to obtain a well-defined,
i.e. time-independent, mass formula without requiring that
$\f$ and $\r$ asymptote to the LDV that fast
as $\stp$, hence allowing
more general coordinate systems like in \iv.

In 4D general relativity there is a very straightforward
method to obtain the total energy (ADM mass) of an
asymptotically Minkowskian configuration. As noticed by
Regge and Teitelboim
\REF\RT{T. Regge and C. Teitelboim, Ann. Phys. {\bf 88}
(1974) 286.}
almost twenty years ago [\RT], in order to have a
well-defined variational principle, i.e. in order that
$$\d\left( \int {\rm d}^3 x \sum_i \dot\varphi_i \Pi_i
-H\right) =0
\eqn\vi$$
for all solutions of the equations of motion with the
required asymptotics, the Hamiltonian $H$ cannot simply be
the bulk Hamiltonian $H_1=\int{\rm d}^3 x {\cal H}$. One
needs to include a boundary hamiltonian
$H_2=\lim_{r\to\infty} \int{\rm d}^2S {\cal M}$ whose
variation cancels (using the boundary conditions) the
boundary terms obtained when varying $H_1$. Since $H_1$ is
a sum of constraints its actual value vanishes for any
solution and the total energy is simply the numerical value
of $H_2$.

We will apply this method here to the 2D dilaton gravity
theories, i.e. to the classical theory \i\ and to the
``quantum versions" of refs. \CGHS\ and \RST. For the latter
theories we represent the trace anomaly by the $Z$-field, as
extensively discussed in refs.
\REF\ST{L. Susskind and L. Thorlacius,
\NP B382 1992 123 .}
\REF\BIL{A. Bilal, ``Positive energy theorem and
supersymmetry in exactly solvable quantum-corrected 2D
dilaton-gravity, Princeton University preprint PUPT-1373,
hep-th@xxx/9301021.}
 [\ST,\PS,\BIL]. We can treat all three models
simultaneously by writing
$$
S = {1\over 4\pi} \is \rg
\left[ \left( e^{-2\f} -{\k\over 2}\f\right) R
+4 e^{-2\f} \left( (\nabla\f)^2 + \l^2\right)
-\half (\nabla Z)^2 +QRZ\right]
\eqn\vii$$
where $\k=Q=0$ gives back the classical action (with one
free matter $Z$-field\foot{
One might add other free (classical) matter fields. These
could be included trivially into  our
subsequent analysis, in particular their contribution to
the boundary term $D$ would vanish due to the standard
boundary conditions on matter fields.}), $\k=0,\ 2Q^2={N\over
12}$ gives the CGHS model [\CGHS], and $\k=2Q^2={N-24\over
12}$ gives the RST-model [\RST]. We will obtain the bulk
Hamiltonian $H_1$ for these theories and then identify the
correct $H_2$ as described above. As expected, $H_1$ is just
a sum of constraints that satisfy (two copies of) a Poisson
bracket Virasoro algebra (or rather stress-energy tensor
algebra). $H_2$ is identified with the mass functional. Its
numerical value coincides with that of the total Hamiltonian
and thus must be constant. We will check that this is indeed
the case.

{ \chapter{The canonical structure}}

First we will derive the bulk Hamiltonian $H_1$ from the
action \vii.
For the classical action only, this was done in
\REF\MIK{A. Mikovic, ``Exactly solvable models of 2d dilaton
gravity, Queen Mary preprint QMW/PH/92/12, hep-th@xxx/9207006,
``Two-dimensional dilaton gravity in a unitary gauge, QMW/PH/92/16,
hep-th@xxx/9211082.}
ref. \MIK.\foot{
This reference, however, does not address the problem of finding
the boundary Hamiltonian which is our main concern here.}
To begin with, we parametrize the
two-dimensional metric in the following way
$$g_{\m\n}=e^{2\r}\ \pmatrix{ A^2-B^2 & A\cr A & 1\cr}\ .
\eqn\di$$
This is inspired by the standard ADM parametrization
\REF\ADM{R.L. Arnowitt, S. Deser and C.W. Misner, in
``Gravitation: An introduction to current research", ed. L.
Witten, Wiley, New York, 1962;
see also: C.W. Misner, K.S. Thorne and J.A. Wheeler,
``Gravitation", W.H. Freeman \& Company, San Francisco,
1973. }
[\ADM] with $A$ and $e^\r B$ the analogues of the
shift vector and lapse function. Due to the Weyl invariance
of the classical action it is more convenient to use $B$ as
a field rather than the lapse function itself. Then
conformal gauge simply is $A=0,\ B=1$. The simplest way to
compute the curvature  probably is to use the zweibein
$e^0=\exp (\r) B {\rm d}t,\ e^1=\exp (\r) (A{\rm d}t+{\rm
d}x)$, compute the spin-connection $\o$ from the zero-torsion
equation, ${\rm d}e^a+\o^{a}_{\phantom{a}b}e^b=0$,
 and obtain ${\rm d}t{\rm d}x\rg R=2{\rm d}\o$. It is then
straightforward to obtain
$(\o^{01})_1={1\over B}(\dot\r-A\r'-A')$ and
$(\o^{01})_0=B\r'+B'+{A\over B}(\dot\r-A\r'-A')$ and hence
$$\eqalign{
S= \is {1\over \pi B}\Bigg\{
&\left[ F(\dot\f-A\f')-{1\over 2}Q(\dot Z-AZ')\right]
(\dot\r-A\r'-A')\cr
-&(F\f'-{1\over 2}QZ')(B^2\r'+BB')\cr
-&e^{-2\f}(\dot\f-A\f')^2+B^2e^{-2\f}\f'^2+
\l^2B^2e^{-2\f+2\r}\cr
+&{1\over 8}(\dot Z-AZ')^2-{1\over 8}B^2Z'^2
\Bigg\}\cr
}
\eqn\dii$$
where $F=e^{-2\f}+{\k\over 4}$. To obtain \dii\ we have
integrated by parts. We will not keep track of boundary
terms for the moment since first we only want to obtain the
bulk Hamiltonian $H_1$. The canonical momenta then are
$$\eqalign{
\pf&={1\over \pi B}\left[ F (\dot\r-A\r'-A')
-2e^{-2\f}(\dot\f-A\f')\right]\cr
\pr&={1\over \pi B}\left[ F (\dot\f-A\f')
-{1\over 2}Q(\dot Z-AZ')\right]\cr
\pz&={1\over \pi B}\left[ {1\over 4}(\dot Z-AZ')
-{1\over 2}Q (\dot\r-A\r'-A') \right]\ . \cr
}
\eqn\diii$$
Since no time-derivatives of $A$ or $B$ occur in the action
there are no momenta conjugate to $A$ or $B$. The fields
$A$ and $B$ are Lagrange multipliers serving to impose
constraints. Writing $S=\int {\rm d}\t L$, the bulk
Hamiltonian is given by $H_1=\int \ds (\dot\f\pf+\dot\r\pr+
\dot Z\pz )-L$ which after integrating by parts reads
$$\eqalign{
H_1=&\int \ds [ A\ca +B\cb ]\cr
\ca=&\f'\pf+\r'\pr+Z'\pz -\pr'\cr
\cb=&{\pi\over G^2}\left[
e^{-2\f}(\pr+2Q\pz)^2+F\pf(\pr+2Q\pz)+{1\over
2}Q^2\pf^2\right] +2\pi\pz^2\cr
&+{1\over \pi}\left[ F\r'\f'-(F\f')'-e^{-2\f}\f'^2-
\l^2e^{-2\f+2\r}+{1\over 8}Z'^2-{1\over 2}Q\r'Z'+
{1\over 2}QZ''\right]}
\eqn\div$$
where $G^2=F^2-2Q^2e^{-2\f}$.

The Lagrange multipliers $A$ and $B$ impose the constraints
$\ca=\cb=0$. To see what these constraints are it is
convenient to substitute \diii\ for the momenta. Then in
conformal gauge ($A=0, B=1$) one has (substituting also
$Z=\zt+2Q\r$) %
$$\eqalign{
\pi\ca\Big\vert_{A=0,B=1}&=F(\dot\r\f'+\r'\dot\f-\dot\f')
-Q^2(\dot\r\r'-\dot\r')+{1\over 4}\dot\zt\zt'+{1\over
2}Q\dot\zt'\cr
\pi\cb\Big\vert_{A=0,B=1}&=F(\dot\r\dot\f+\r'\f'-\f'')
-e^{-2\f}(\dot\f^2-\f'^2)-{1\over
2}Q^2(\dot\r^2+\r'^2-2\r'')\cr
&-\l^2e^{-2\f+2\r}+{1\over 8}(\dot\zt^2+\zt'^2+4Q\zt'') \cr
}
\eqn\dv$$
which coincides with ${1\over 2}(T_{++}-T_{--})$ and
${1\over 2}(T_{++}+T_{--})$, respectively, of refs. \CGHS,
\RST,\foot{
More precisely, the $T_{\pm\pm}$ of refs. \CGHS,
\RST\ contain $\pa_{\pm}^2({\rm fields})$ which includes
$\pa_{\t}^2({\rm fields})$. In a canonical formalism this
must be replaced by $\pa_{\s}^2({\rm fields}) + \ldots $
using the equations of motion. Once we do this,
${1\over 2}(T_{++}\pm T_{--})$,  of refs. \CGHS,
\RST\ are identical to the above $\pi\ca, \pi\cb$ in
conformal gauge. In fact, the equation of motion one has to
use coincides with $T_{+-}=0$ and one finds that $\pi\cb$ as
given by \div\ is precisely
${1\over 2}(T_{++}+T_{--}) +T_{+-}=2T_{00}$
while $\pi\ca$ is ${1\over 2}(T_{++}-T_{--})=2T_{01}$
as expected on general grounds.
}
the $\zt$-part corresponding to the matter (or matter plus
ghost) stress tensor.

Using canonical Poisson brackets,
$$\{\f(\s),\pf(\s')\}=
\{\r(\s),\pr(\s')\}=
\{Z(\s),\pz(\s')\}=\d (\s-\s')
\eqn\dvi$$
we can compute the algebra of the constraints as given by
\div\ (in general gauge). It is straightforward to obtain:
$$\eqalign{
\{\ca(\s),\ca(\s')\}&=(\pa_\s-\pa_{\s'})[\ca(\s')\d(\s-\s')]
\ , \cr
\{\cb(\s),\cb(\s')\}&=(\pa_\s-\pa_{\s'})[\ca(\s')\d(\s-\s')]
\ , \cr
\{\cb(\s),\ca(\s')\}&=(\pa_\s-\pa_{\s'})[\cb(\s')\d(\s-\s')]
\ . \cr}
\eqn\dvii$$
Thus we see that the Poisson bracket of $\ca+\cb$ with
$\ca-\cb$ vanishes while
$$
\{ (\cb(\s)\pm \ca(\s)),(\cb(\s')\pm \ca(\s')) \}
=2(\pa_\s-\pa_{\s'})[(\cb(\s')\pm \ca(\s'))\d(\pm\s\mp\s')]
\eqn\dviii$$
which is indeed the Poisson bracket algebra of $T_{\pm\pm}$
with itself. (Note that for $T_{--}$ we have $\d(-\s+\s')$
on the r.h.s. since $T_{--}$ naturally is a function of
$\t$ and $-\s$.) There is no $\d'''$-term which means that
the total central charge vanishes. This is indeed the case
as one can easily see in conformal gauge: For the classical
theory this is obvious. For the CGHS-model the conformal
anomaly term $\sim Q^2(\pa_\pm\r\pa_\pm\r -\pa_\pm^2\r)$
gives $c=-24Q^2=-N$ while the matter fields represented by
$\zt$ give $c=+N$. For the RST-model  $Q^2(\pa_\pm\r\pa_\pm\r
-\pa_\pm^2\r)$ gives $c=-24Q^2=-12\k=24-N$, while the
$\zt$-field gives the anomaly for matter, ghosts and the
quantum part of $\f,\r$ which is $c=N-26+2=N-24$. Of
course, we just repeated that the Polyakov-anomaly action
is designed to cancel the various anomalies present in the
theory.

Let us now compute the variation of the bulk Hamiltonian
$H_1$ as given by \div\ under infinitesimal variations of
the physical fields and their
canonical momenta, this time keeping track of all boundary
terms:
$$\eqalign{
\d H_1=&\int \ds\Big[
{\d H_1\over \d\f} \d \f
+{\d H_1\over \d\r} \d \r
+{\d H_1\over \d Z} \d Z
+{\d H_1\over \d\pf} \d \pf
+{\d H_1\over \d\pr} \d \pr
 +{\d H_1\over \d\pz} \d \pz \Big]\cr
  &+D }
\eqn\dix$$
where $D$ is the boundary term given below and
$$\eqalign{
{\d H_1\over \d\f}=&{2\pi B\over G^4}e^{-2\f}\Big[
F(e^{-2\f}-{\k\over 4})
(\pr+2Q\pz)^2\cr
&\phantom{
{2\pi B\over G^4}e^{-2\f}
}
+(F^2-{1\over 2}Q^2\k )\pf(\pr+2Q\pz)
+(F-Q^2)Q^2\pf^2\Big]\cr
&+ {1\over
\pi}\left[-(BF\r')'-B''F+2Be^{-2\f}\f'^2+(2Be^{-2\f}\f')'-
2B e^{-2\f}\r'\f'+2\l^2B e^{-2\f+2\r}\right]\cr &-(A\pf)'\cr
{\d H_1\over \d\r}= &{1\over \pi}\left[-(BF\f')' +{1\over
2}Q(BZ')'-2\l^2B e^{-2\f+2\r}\right] -(A\pr)'\cr
{\d H_1\over \d Z}= &{1\over 2\pi}\left[QB'' +Q(B\r')'
-{1\over 2}(BZ')'\right] -(A\pz)'\cr
 }
\eqn\dx$$ %
and
$$\eqalign{
{\d H_1\over \d\pf}=&{\pi B\over G^2}\left[
F(\pr+2Q\pz) +Q^2\pf\right] +A\f'\cr
{\d H_1\over \d\pr}=&{\pi B\over G^2}\left[
2e^{-2\f}(\pr+2Q\pz) +F\pf\right] +A\r'+A'\cr
{\d H_1\over \d\pz}=&4\pi B\pz+{2\pi Q B\over G^2}\left[
2e^{-2\f}(\pr+2Q\pz) +F\pf\right] +AZ'\cr
}
\eqn\dxi$$
Hamilton's equations, ${\d H_1\over
\d\varphi_i}+\dot\Pi_i={\d H_1\over
\d\Pi_i}-\dot\varphi_i=0$, would follow from the
variational principle $\d \left( \int\ds
(\dot\f\pf+\dot\r\pr+\dot Z\pz) - H_1\right)=0$ if the
boundary term $D$ would vanish. We have (recall that
$F=e^{-2\f}+{\k\over 4}$)
$$\eqalign{ \pi D = \Bigg[
&\d\left(-BF\f'+{1\over 2}QBZ'\right) +
\left( B'F +BF\r'-2Be^{-2\f}\f'+A\pf\right)\d\f\cr
&+\left( BF\f'-{1\over 2}QBZ'+A\pr\right)\d\r
+\left( -{1\over 2}QB'-{1\over 2}QB\r'+{1\over 4}BZ'+A\pz
\right)\d Z\cr
&-A\d\pr \Bigg]^{\sep}_{\sem}\ .\cr}
\eqn\dxii$$
This does not vanish. In the next section, we will show,
however, that, subject to appropriate boundary conditions,
$D$ can be written as the variation of another functional
$H_2$:
$$ D\Big\vert_{\rm boundary\ conditions} =-\d H_2\ .
\eqn\dxiii$$
Then
$$\d \left( \int\ds
(\dot\f\pf+\dot\r\pr+\dot Z\pz) - H_1-H_2\right)=0
\quad \Rightarrow \quad {\rm Hamilton's\ equations}\ .
\eqn\dxiv$$
Thus the true Hamiltonian $H$ is the bulk Hamiltonian $H_1$
plus the boundary Hamiltonian $H_2$. Since $H_1$ is only
given by the sum of the two constraints it vanishes on all
solutions and the total energy, i.e. the value of the total
Hamiltonian $H$, is given by the value of $H_2$ only.

{\chapter{ The total energy for asymptotically Minkowskian
space-times}}

Obviously, in general, $D$ cannot be written as the
variation of some functional. Indeed, we know that the
notion of total energy is well-defined only if we impose
appropriate boundary conditions. In 4D general relativity
one not only requires the space-time to be asymptotically
flat but also to be asymptotically Minkowskian, i.e. one
imposes asymptotic coordinate conditions. Here we will
require that asymptotically, as $\s\to\pm\infty$, we have
the LDV: $\f\sim -\l\s,\ \r\sim 0$.
Hence we impose the following asymptotics as $\stm$:
$${\rm as}\ \stm\ : A\sim 0,\ B\sim1,\ \r\sim 0,\
\f\sim-\l\s,\ Z\sim 0\ ,
\eqn\ti$$
and it is understood that the derivatives of the fields
obey the derivatives of these relations\foot{
Although this is obvious for ``smooth" field
configurations, it has to be imposed separately to exclude
configurations with asymptotics like $\f\sim -\l\s +
e^{\l\s}\sin (e^{-2\l\s})$ as $\stm$.}, e.g. $\f'\sim -\l$,
etc. With \diii\ in mind we also require
$${\rm as}\ \stm\ :  \pf\sim 0,\ \pr\sim 0,\ \pz\sim 0\ .
\eqn\tii$$
These asymptotic conditions then imply that $D$ receives no
contribution from $\sem$.

For $\stp$, however, just as in 4D general
relativity [\RT], we have to be more precise about how fast
the field configurations actually have to approach the LDV.
The asymptotic conditions to be imposed tor $\stp$ should
be satisfied for all solutions of the equations of motion
and constraints. We will suppose that the matter fields
(here represented by $Z-2Q\r$) asymptotically vanish
(excluding radiation baths that lead to infinite total
energy). Then it follows from ref. \CGHS\ for the classical
case in conformal gauge that as $\stp$ we have
$\f\sim-\l\s+(\alpha e^{\l\t}+\beta e^{-\l\t}) e^{-\l\s}
+\ldots$ and (in an appropriate class of coordinates) also $\r\sim
(\alpha e^{\l\t}+\beta e^{-\l\t}) e^{-\l\s} +\ldots$.
For the quantum models, differences with the classical
model only appear ${\cal O}(e^{2\f})\sim {\cal O}(e^{-2\l\s})$, thus
in all cases the deviations from the LDV are at least ${\cal
O}(e^{-\l\s})$. It is sufficient if the phase-space includes
only fields with these asymptotics. Hence we require
$$\eqalign{
{\rm as}\ \stp\ :\quad \f+\l\s,\ \r ,\ B-1   &\sim e^{-\l\s}\cr
\pf,\ \pr &\sim e^{\l\s}\cr
Z,\ \pz &\to 0\cr
e^{\l\s} A &\to 0\ .\cr}
\eqn\tiii$$
This is meant as a minimal requirement, i.e. $\r$
and $\f+\l\s$ decrease as $e^{-\l\s}$ or faster, and $\pf,
\pr$ do not grow faster than $e^{\l\s}$. Finally we also
require %
$$
{\rm as}\ \stp\ :\quad e^{2\l\s} (\f+\l\s-\r)\to 0\ .
\eqn\tiv$$
Note that we do {\it not} require $\f+\l\s$ or $\r$ to be
${\cal O}(e^{-2\l\s})$, only the combined relation \tiv\
should be satisfied. Note also that we could have dropped the
$\r$-asymptotics from \tiii\ since it is implied by \tiv\
and the $\f$-asymptotics. Condition \tiv\ specifies a
certain class of coordinate systems.
We know, for the classical model and for RST, that (in
conformal gauge) $\f-\r$ is a free field. Then by a
(residual) coordinate choice we can achieve $\f=-\l\s+\r$.
For CGHS this receives higher-order corrections. This is
the motivation for our boundary condition \tiv.  We will
have more to say about the meaning of it below.  Since all
fields in the phase space satisfy \tiii, \tiv\ the same
applies to the variations $\d\f, \d\r, \d Z, \d\pf, \d\pr$
and $\d \pz$, in particular, we can substitute $\d\r=\d\f$ in
\dxii.
Again it is understood that the derivatives of the fields
obey the derivatives of these relations.
 Using the boundary conditions \tiii\ it is easy to see
that the only terms in $D$ (eq. \dxii) contributing in the
$\stp$-limit are those proportional to $e^{-2\f}$. All
others vanish in this limit. Then we have
$$\pi D= \left[-\d \left( e^{-2\f} B\f' \right)
+e^{-2\f}\left( B'+B\r'-2B\f'\right) \d\f
+e^{-2\f} B\f'  \d\r\right]
 \Bigg\vert_{\sep}
\eqn\tva$$
Using \tiv\ one gets
 $$\eqalign{
\pi D&= \left[-\d \left( e^{-2\f} B\f' \right)
+e^{-2\f}\left( B'+B\l\right) \d\f \right]
 \Bigg\vert_{\sep}\cr
&= -\d \left[ e^{-2\f}\left( B\f' +{1\over 2} B\l
+{1\over 2} B'\right) \right] \Bigg\vert_{\sep} }
\eqn\tv$$
and we identify the boundary Hamiltonian $H_2$ as
$$H_2= {1\over 2\pi} \left[ e^{-2\f}\left( 2 B\f' +
B\l + B'\right) +\l e^{2\l\s} \right] \Bigg\vert_{\sep}
\eqn\tvi$$
where we adjusted an (infinite) additive field-independent
term, not affecting the relation $\d H_2=-D$, so that $H_2$
vanishes for the LDV. The total energy is simply given by
$M=2\pi H_2$:
$$M=  \left[ e^{-2\f}\left( 2 B\f' +
B\l + B'\right) +\l e^{2\l\s} \right] \Bigg\vert_{\sep}
\eqn\tvii$$
This is the general mass formula for all configurations
subject to \ti-\tiv. If we are in conformal gauge ($B=1$)
and write $\f\sim -\l\s+\d\f$ with $\d\f$ at least ${\cal
O}(e^{-\l\s})$ then
$$M=  \left[ 2e^{2\l\s} (\pa_\s +\l) (\d\f -\d\f^2) \right]
\Bigg\vert_{\sep}
\eqn\tviii$$
The old mass formula \iii\ (with our boundary conditions)
only is the part linear in $\d\f$. If we evaluate $M$ as
given by \tviii\ on the solution \iv\ we correctly find
$M=a$, thanks to the $\d\f^2$-term.

More generally, it is
straightforward to check using the
constraints and boundary conditions that $\dot M =0$.
Indeed, we have from \tvii, using \tiv\ (which allows us to
replace $\dot\f$ by $\dot\r$, etc.)
$$
\dot M=  -2  e^{-2\f} \left[ B \left( \f'\dot\r +
\r'\dot\f-\dot\f' \right) + B'\dot\f \right]
\Bigg\vert_{\sep} \ .
\eqn\tix$$
On the other hand, inserting the definitions of the momenta
\diii\ into the constraint $\ca$ \div\ and using the
boundary conditions \tiii, \tiv\ we obtain as $\stp$
$$ \ca \sim   {e^{-2\f}\over \pi B}   \left( \f'\dot\r +
\r'\dot\f-\dot\f'  + {B'\over B} \dot\f \right)\ .
\eqn\tx$$
Thus we conclude
$$ \dot M= -2 \pi B^2 \ca\Bigg\vert_{\sep} \ ,
\eqn\txi$$
and since $\ca=0$ is a constraint, $\dot M$ vanishes. Let
us stress that $M$ is a constant by virtue only of the
constraints and boundary conditions, independently of
whether or not the equations of motion are satisfied.
Note that the same mass formula \tvii\ and \tviii\ apply to
all three models, classical, CGHS and RST.

Finally, we would like to comment on the condition \tiv. It
should be considered as an asymptotic coordinate condition.
That such a condition is necessary is rather obvious: In
ordinary general relativity the vacuum is translationally
invariant and we can make asymptotic Lorentz
transformations. In dilaton-gravity, however, the vacuum is
the LDV which is not translationally invariant. Thus only
the total energy is a meaningful concept, provided it is
defined with respect to the time-like Killing vector
orthogonal to the vector singled out by the LDV. Thus the
asymptotic coordinates have to be chosen carefully, which
is achieved by \tiv. It is also worth pointing out that we
can weaken some of the boundary conditions. Here, we will
not explore this much further. We only note that if one works
in conformal gauge, $A=0, B=1$, \tiii\ can be replaced by
$\f+\l\s, \r, Z \to 0$, as $\stp$, i.e. we only require
LDV asymptotics (without saying how fast the LDV is
approached), and \tiv. Then we still obtain the same
boundary Hamiltonian and total energy \tvii\ (with $B=1,
B'=0$).

{\chapter{Conclusions}}

We have worked out the canonical structure of 2D dilaton
gravity theories (classical, CGHS and RST). As in 4D
general relativity, the bulk Hamiltonian alone does not
lead to a well-defined variational principle. After defining
the phase space carefully by chosing appropriate boundary
conditions we were able to render the variational principle
well-defined by adding a boundary Hamiltonian to the bulk
one. Since the bulk Hamiltonian vanishes by the constraints
the total energy is given by the boundary Hamiltonian
alone. This total energy must be conserved, a fact we also
checked directly.

\ack

It is a pleasure to acknowledge stimulating
discussions with Curt Callan and
Larus Thorlacius. I.K. is supported by the National Science
Foundation grant NSF PHY90-21984.

\refout

\end